\documentclass{PoS}
\usepackage{xspace}
\usepackage[utf8]{inputenc}
\RequirePackage{amsmath}
\RequirePackage{amssymb}
\RequirePackage{mathtools}
\RequirePackage{slashed}
\RequirePackage{graphicx}
\RequirePackage{array}
\RequirePackage{enumerate}
\RequirePackage{xcolor}
\RequirePackage[utf8]{inputenc}
\RequirePackage[british]{babel}
\RequirePackage{cite}
\RequirePackage{booktabs}
\RequirePackage{multirow}
\RequirePackage{url}
\RequirePackage{xargs}
\RequirePackage{pifont}
\DeclareRobustCommand{\ensuremathrm}[1]{\ensuremath{\mathrm{#1}}\xspace}

\DeclareRobustCommand{\NNLOJET}{\textsc{NNLOjet}}
\DeclareRobustCommand{\NNPDF}{\texttt{NNPDF31\_nnlo}}
\DeclareRobustCommand{\rT}{\ensuremathrm{T}}
\DeclareRobustCommand{\ptcut}{{\ensuremath{p_{\rT,\mathrm{cut}}^W}}\xspace}
\DeclareRobustCommand{\PZ}{{\ensuremathrm{Z}}\xspace}
\DeclareRobustCommand{\PW}{{\ensuremathrm{W}}\xspace}
\DeclareRobustCommand{\PWJ}{{\ensuremathrm{W+jet}}\xspace}
\DeclareRobustCommand{\PWp}{{\ensuremathrm{W^+}}\xspace}
\DeclareRobustCommand{\PWm}{{\ensuremathrm{W^-}}\xspace}
\DeclareRobustCommand{\PWpm}{{\ensuremathrm{W^\pm}}\xspace}

\DeclareRobustCommand{\ptw}{{\ensuremath{p_\rT^\PW}}\xspace}

\DeclareRobustCommand{\GeV}{\ensuremathrm{GeV}\xspace}
\DeclareRobustCommand{\TeV}{\ensuremathrm{TeV}\xspace}

\title{NNLO QCD Corrections to W+jet Production in \NNLOJET}

\ShortTitle{NNLO QCD Corrections to W+jet Production in \NNLOJET}
\author{Aude Gehrmann-De Ridder\\
  Institute for Theoretical Physics, ETH, CH-8093 Zürich, Switzerland\\
  Department of Physics, University of Zürich, CH-8057 Zürich, Switzerland\\
        E-mail: \email{gehra@phys.ethz.ch}}
\author{Thomas Gehrmann\\
  Department of Physics, University of Zürich, CH-8057 Zürich, Switzerland\\
        E-mail: \email{thomas.gehrmann@uzh.ch}}
\author{Nigel Glover\\
        Institute for Particle Physics Phenomenology, Department of Physics, University of Durham,
Durham, DH1 3LE, UK\\
        E-mail: \email{e.w.n.glover@durham.ac.uk}}
\author{Alexander Huss\\
  Theoretical Physics Department, CERN, 1211 Geneva 23, Switzerland\\
        E-mail: \email{alexander.huss@cern.ch}}
\author{\speaker{Duncan Walker}\\
        Institute for Particle Physics Phenomenology, Department of Physics, University of Durham,
Durham, DH1 3LE, UK\\
        E-mail: \email{duncan.m.walker@durham.ac.uk}}

\abstract{
  We give an overview of our calculation of the next-to-next-to-leading order (NNLO) QCD corrections to \PW + jet production in hadronic collisions. Phenomenological results for multiple differential distributions are compared to CMS data for 8 \TeV proton-–proton collisions. We further discuss the application of the calculation to the transverse momentum spectrum of inclusive \PW boson production, again accompanied by a comparison to 8 \TeV CMS data. In both cases, the inclusion of NNLO QCD effects give an improved agreement between theory and data with considerably reduced scale uncertainties with respect to the next-to-leading order (NLO) results.
}

\FullConference{Loops and Legs in Quantum Field Theory (LL2018)\\
		29 April 2018 - 04 May 2018\\
		St. Goar, Germany}
\begin{document}

\section{Introduction}

The production of \PW bosons in hadron-hadron collisions is one of the most important processes in terms of our understanding of the Standard Model. \PW bosons decaying leptonically have a large experimental cross section, characterised by a lepton with high transverse momentum accompanied by missing energy in events corresponding to the produced (anti-)neutrino. Knowledge of the process to a high precision in both theory and experiment is mandatory, informing our understanding of subjects including detector calibration and the accuracy of calculational techniques over a large kinematic range.

Many important measurements can be made in \PW production. In particular, W production in association with a jet is a dominant background for much rarer processes including associated WH production and single top quark production. It is also crucial for dark matter searches with a missing energy signal, where the process is an important background. Hadronic W production also gives us a valuable insight into the flavour content of parton distribution functions (PDFs), principally through the different couplings of the $\PWpm$ bosons to valence quarks inside the proton. Measurements of \PW production with a single associated jet also allow information to be obtained with respect to the proton gluon content, which appears at leading order (LO) for the process.

Further to this, calculations for \PWJ production can also be utilised for inclusive production at finite transverse momentum \ptw, where a recoiling jet is implicitly required for momentum conservation. Accurate predictions for \ptw are particularly important for \PW mass determinations where the $p_T$ spectrum of the \PZ boson is used to indirectly model and estimate behaviours in the \ptw spectrum. The large size of the \PW production cross sections ensures small statistical errors in experimental measurements which offsets the systematic errors coming from $E_T$ reconstruction for the neutrino. % Link in to low statistical errors

Naturally, due to the importance of the process, considerable work has been done with regard to improving the theoretical predictions. In terms of electro-weak (EW) calculations, next-to-leading order (NLO) corrections to the inclusive process have been computed in \cite{Denner:2009gj,Kallweit:2015dum}. For the QCD corrections, NLO + parton shower (PS) has become standard for low jet multiplicities and is now automated in multiple general-purpose event generators. The next-to-next-to-leading order (NNLO) QCD corrections to inclusive \PW production have been known for a long time \cite{Hamberg:1990np,Anastasiou:2003ds,Melnikov:2006di}, and have been independently reproduced in multiple calculations. The NNLO QCD corrections for W production in association with a single jet,
\begin{equation}
 p p \rightarrow \PWpm(\rightarrow l+\nu_l) + \mathrm{jet}+X,
\end{equation}
were calculated much more recently \cite{Boughezal:2016dtm,Boughezal:2015dva,Gehrmann-DeRidder:2017mvr}, and it is the computations of these corrections with the \NNLOJET~ framework which we consider here.

\NNLOJET~is a Monte Carlo parton-level event generator at NNLO in QCD  using the method of antenna subtraction \cite{GehrmannDeRidder:2005cm,GehrmannDeRidder:2005aw,GehrmannDeRidder:2005hi,Daleo:2006xa,Daleo:2009yj,Boughezal:2010mc,Gehrmann:2011wi,GehrmannDeRidder:2012ja,Currie:2013vh} for the cancellation of the infra-red divergences between real and virtual terms in the perturbative expansion in the strong coupling $\alpha_S$. The subtraction counterterms are formed from ratios of matrix elements for simple processes which can be used to fully replicate the divergent structures of more complex processes. All relevant real emission antenna functions for all hadron-hadron collisions with massless partons have been integrated over the unresolved phase space to give the virtual counterparts, meaning that all components required for NNLO subtraction in massless QCD are well known and allowing computation of cross sections and multidifferential distributions with arbitrary cuts in \NNLOJET. 

Following results for boson production in association with a jet~\cite{Ridder:2015dxa,Ridder:2016nkl,Gehrmann-DeRidder:2016jns,Gauld:2017tww,Chen:2018pzu,Chen:2016zka,Chen:2014gva,Gehrmann-DeRidder:2017mvr} and di-jet production~\cite{Currie:2017eqf,Currie:2016bfm,Currie:2018xkj} in proton-proton collisions, di-jet production in neutral current, charged current and diffractive DIS~\cite{Niehues:2018was,Currie:2016ytq,Currie:2017tpe,Britzger:2018zvv} and three-jet production in $e^+ e^-$-annihilation~\cite{Gehrmann:2017xfb}, the process library in \NNLOJET~has recently been expanded to include Higgs production in vector boson fusion (VBF) in proton-proton collisions~\cite{Cruz-Martinez:2018rod}, and single-jet production to N3LO QCD in NC DIS~\cite{Currie:2018fgr}, using the method of Projection-To-Born (P2B).

In Section \ref{sec:CMS_8TeV_standard} we present new results for CMS 8TeV data for W production in association with a jet, and in Section \ref{sec:CMS_8TeV_ptw} we show results for the closely related $p_T$ spectrum of the \PW, first reported in \cite{Gehrmann-DeRidder:2017mvr}.

\section{\label{sec:CMS_8TeV_standard}\PW boson production in association with a hadronic jet}
% Done with N-jettiness NNLO/SHERPA+Blackhat NLO in the paper -> Mention?
% We have booked the 2J NLO distributions made with SHERPA/Blackhat as well
% A direct comparison can't be done as their results are corrected for hadronisation
% and MPI (and SHERPA includes PDF uncertainties)
% N-jettiness also didn't do the deltaphi distribution

% Correction sizes in each distribution

\begin{figure}[!t]
  \centering
  \begin{tabular}{@{}cc@{}}
  \includegraphics[page=26,scale=0.49]{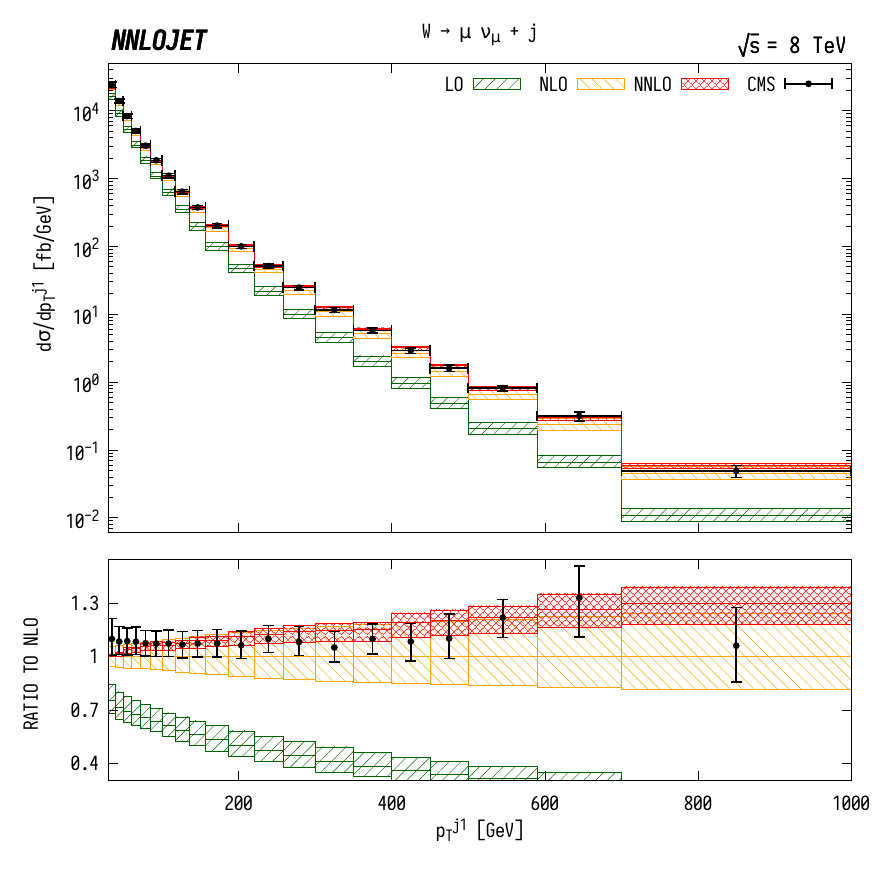} &
  \includegraphics[page=27,scale=0.49]{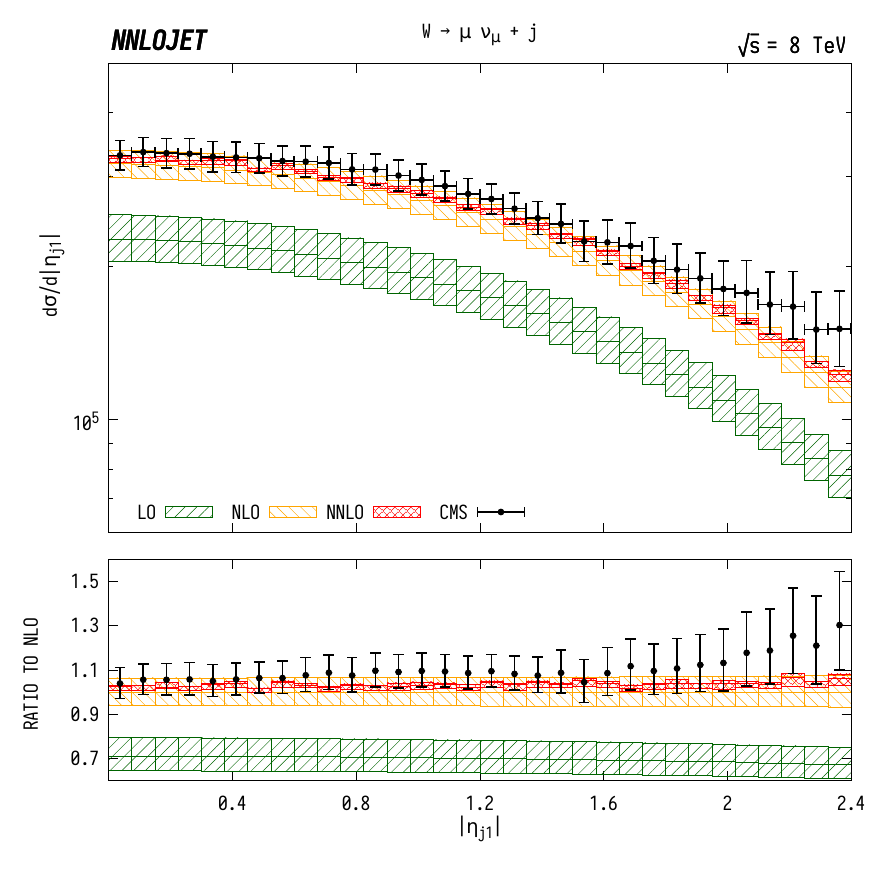} \\
  \end{tabular}
  \caption{Cross sections differential in $|\eta_{j_1}|$, $p_T^{j_1}$ for $\PW$ production compared to CMS data at 8~\TeV. Predictions at LO (green right-hatched), NLO (orange left-hatched), and NNLO (red cross-hatched) are compared to CMS data from  Ref.~\cite{Khachatryan:2016fue}. Distributions normalised to NLO are shown in the lower panels. The bands correspond to scale uncertainties estimated as described in the main text.}
  \label{fig:CMS_8TeV_Standard}
\end{figure}

%% For pure \PWJ production alongside a resolved jet, 
%% Why do we care about W+1J production for these cuts?

We start with results for \PW production, where one or more jets is explicitly required alongside the boson in the experimental analysis. The $19.6~\mathrm{fb}^{-1}$ CMS data taken at 8 \TeV used for this comparison is taken from \cite{Khachatryan:2016fue} for the muonic decay channel in combined \PWpm production, where we use the notation $\PW = \PWp + \PWm$. The $\mathrm{anti-}k_T$ jet algorithm is used, with radius parameter $R=0.5$, and the central scale chosen for the predictions is
\begin{equation}
  \mu_{R}=\mu_{F}=\sqrt{m^2_{\mu\nu}+\sum_i(p^i_{T,\mathrm{jet}})^2},
\end{equation}
with scale variations performed independently for the factorisation and renormalisation scales $\mu_{F}$ and $\mu_{R}$ by factors of $\frac{1}{2}$ and $2$ under the constraint $\frac{1}{2}<\mu_{F}/\mu_{R}<2$. We use the central member of the \NNPDF~ PDF set \cite{Ball:2017nwa} with $\alpha_S(M_Z)=0.118$ for these predictions at all perturbative orders.

The phase space cuts applied for the analysis are:
\begin{align} % Condense to 2 lines?
  p_T^{\mathrm{jets}}>30~\GeV,  \quad |\eta|^{\mathrm{jets}}<2.4, \quad |y^{\mathrm{lep}}|<2.1, \quad p_T^{\mathrm{lep}} > 25~\GeV,  \quad m_T^W > 50~\GeV\mathrm{,}
\end{align}
  where $m_T^W$ is the transverse mass of the $\PW$ boson. Scale errors are treated as fully correlated between the $\PWp$ and the $\PWm$ when obtaining total rates for $\PW$ production. The NNLO QCD distributions were independently calculated for all observables with a single jet at leading order in \cite{Khachatryan:2016fue} using the method of N-jettiness subtraction \cite{Boughezal:2015dva}, however a direct comparison cannot be immediately performed here as hadronisation effects have been applied to the N-jettiness results presented in \cite{Khachatryan:2016fue}.

\begin{figure}[!t]
  \centering
  \begin{tabular}{@{}cc@{}}
  \includegraphics[page=29,scale=0.49]{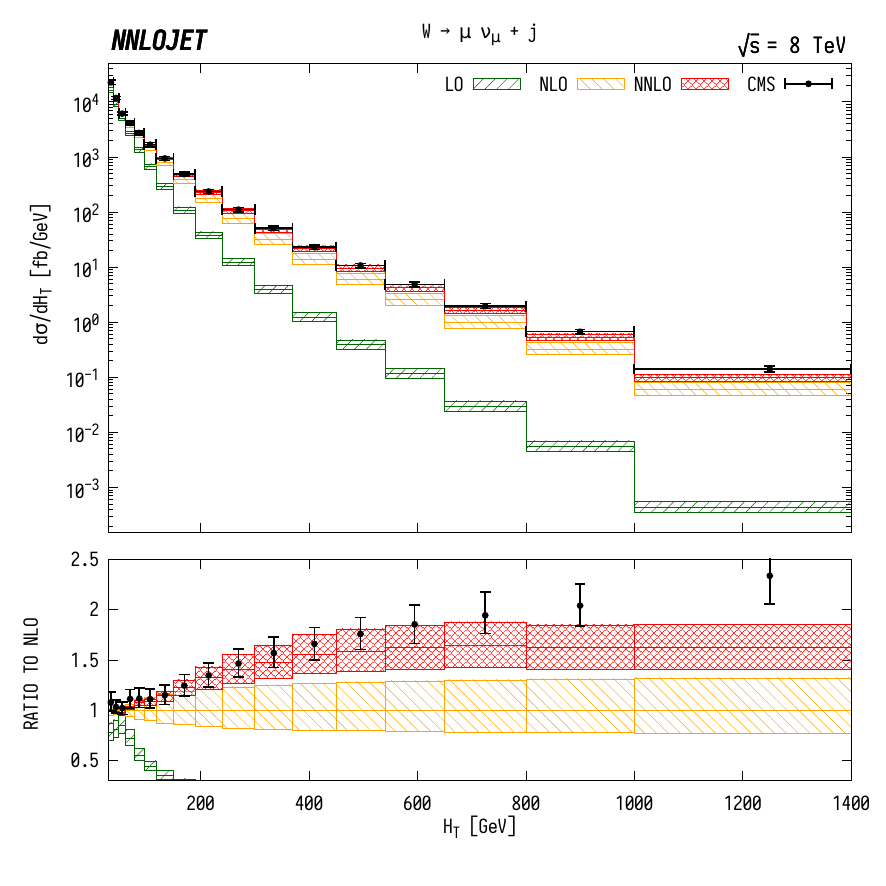} &
  \includegraphics[page=25,scale=0.49]{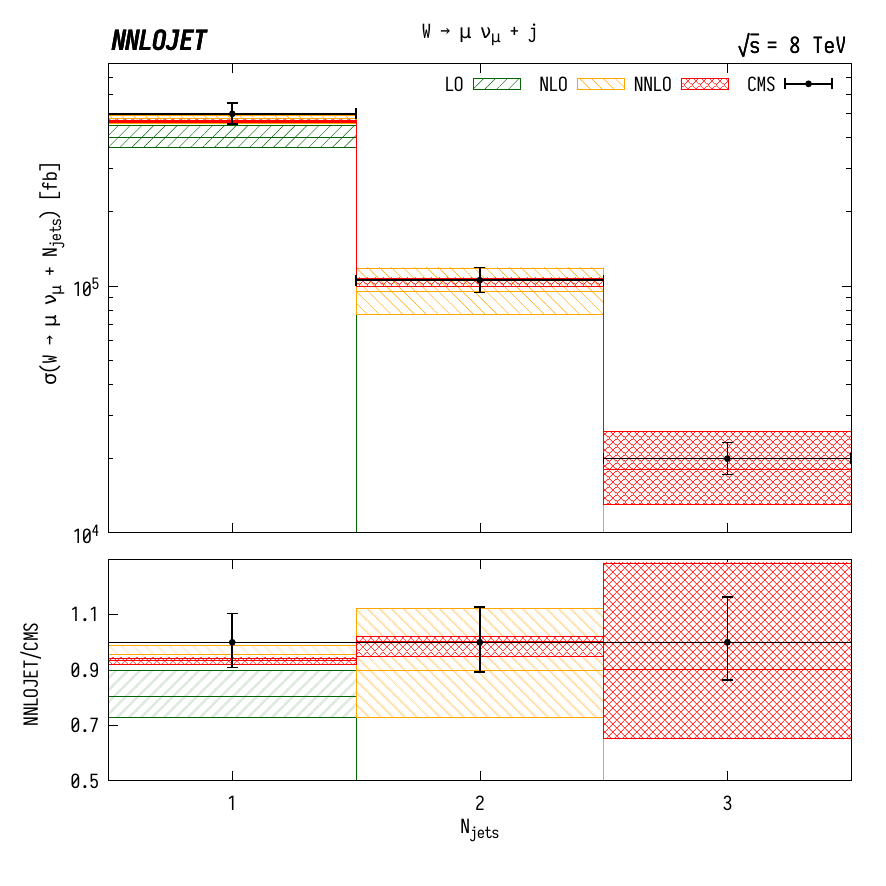} \\
  \end{tabular}
  \caption{The cross section differential in $H_T$ of the jets and the exclusive jet multiplicity $N_\mathrm{jets}$ for $\PW$ production compared to CMS data at 8~\TeV. Predictions at LO (green right-hatched), NLO (orange left-hatched), and NNLO (red cross-hatched) are compared to CMS data from  Ref.~\cite{Khachatryan:2016fue}. The $H_T$ distribution is normalised to NLO, and the jet multiplicity is normalised to data. The bands correspond to scale uncertainties estimated as described in the main text.}
  \label{fig:CMS_8TeV_Standard_2}
\end{figure}

The differential distributions for the absolute pseudorapidity $|\eta_{j_1}|$ and transverse momentum $p_T^{j_1}$ of the leading jet, the scalar sum of jet transverse momenta $H_T$ of the jets and the average number of jets per event $N_{\mathrm{jets}}$ have been computed and are shown in Figs.~{\ref{fig:CMS_8TeV_Standard}} and {\ref{fig:CMS_8TeV_Standard_2}}. For all distributions, we observe a sizeable reduction in scale errors at NNLO with respect to NLO. The $p_T^{j_1}$ distribution shows good agreement with data within errors for the full range of $p_T^{j_1}$ from $30~\GeV$ to $1~\TeV$, with a correction factor of up to 30\% over the NLO result for high $p_T^{j_1}$. There is a shape difference seen at low $p_T$ close to the edge of the allowed phase space where resummation effects begin to have a large impact. The  $|\eta_{j_1}|$ distribution of the leading jet shows good agreement with data up to  $|\eta_{j_1}|\sim 2$ where the fixed order prediction begins to underestimate the data by up to $\sim20\%$, although this is still largely within the experimental error bounds.

The distribution of the scalar sum of jet transverse momenta ($H_T$) shows good agreement with data up to $\sim 700~\GeV$ where effects of higher jet multiplicities $N_{\mathrm{jets}}>3$ than can be described in an NNLO calculation become dominant. This is a marked improvement over NLO, where the agreement of the NLO prediction begins to fail at $\sim200\GeV$. The $N_{\mathrm{jets}}$ distribution shows the exclusive jet multiplicity for the data. There is good agreement for the 1 jet case, and agreement within larger scale variation bands is found for 2 and 3 jets where the NNLO W+J predictions are only NLO and LO accurate respectively.

The origin of the large NNLO/NLO $k$-factors visible at high $p_T^{j_1}$ and $H_T$ can be traced back to events with a dijet type topology, where a relatively soft \PW boson is radiated alongside two hard back-to-back jets \cite{Rubin:2010xp}. These topologies only occur for the first time at NLO in the \PWJ calculation, yet are the dominant contibution for high jet $p_T$. As the full NNLO calculation can only effectively describe these contributions to NLO, we observe sizeable higher order corrections in the regions of phase space where they contribute, accompanied by larger scale variation bands than might otherwise be expected.

\section{\label{sec:CMS_8TeV_ptw}Inclusive \PWpm boson production for finite \ptw}

\begin{figure}[!t]
  \centering
  \begin{tabular}{@{}cc@{}}
\includegraphics[scale=0.85]{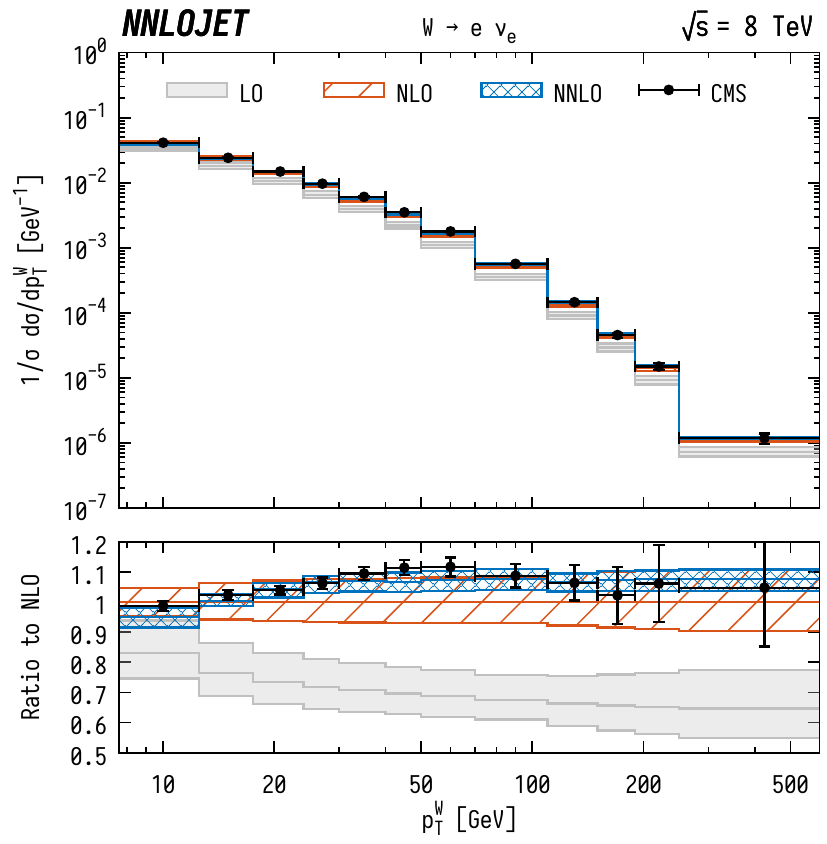} &
\includegraphics[scale=0.85]{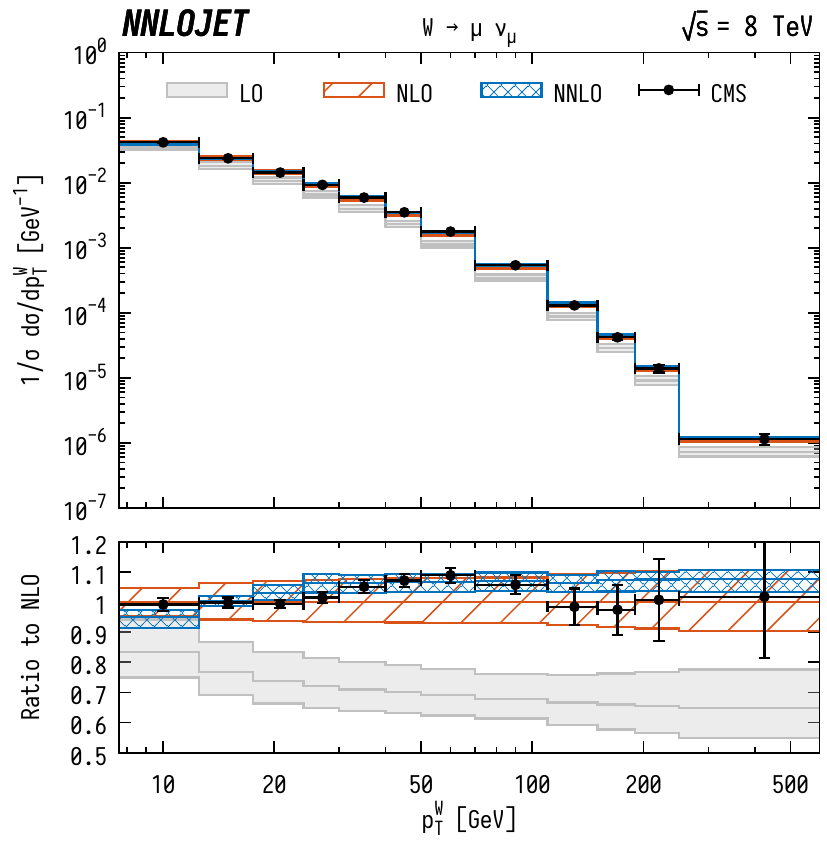}
  \end{tabular}
  \caption{Normalised \ptw distributions for \PW production with subsequent decay into leptons. The left hand panel shows results for electron decays, whereas the right panel shows muonic decays.  Predictions at  LO  (gray  fill),  NLO  (orange  hatched),  and  NNLO  (blue cross-hatched) are compared to CMS data at 8~\TeV from \cite{Khachatryan:2016nbe}. The bands correspond to scale uncertainties estimated as described in the main text.}
  \label{fig:CMS_8TeV_ptw}
\end{figure}

As mentioned previously, our \PWJ calculation can also be used to provide fixed order predictions for the \ptw spectrum in inclusive \PW production above some cut-off \ptcut due to the implicit requirement of a balancing jet:

\begin{equation}
 p p \rightarrow \PWpm(\rightarrow l+\nu_l)\bigr\rvert_{\ptw>\ptcut} +X.
\end{equation}
In comparison to calculations performed to $\mathcal{O}(\alpha_S^2)$ for inclusive \PW production which are trivial at LO, this allows us to extend predictions to $\mathcal{O}(\alpha_S^3)$ for the majority of phase space, excluding the region below the cut-off.
%% By placing a finite cut on the \ptw spectrum,to ensure perturbative stability, one can then directly use a \PWJ calculation to give the fixed order results above this cut. This is important, as state of the art inclusive \PW predictions only extend to $\mathcal{O}(\alpha_S^2)$, and therefore are only NLO accurate for the \ptw spectrum. By using our fixed order calculation, we can extend this to $\mathcal{O}(\alpha_S^3)$, which is particularly important for measurements of the \PW boson mass $M_W$.

We demonstrate this by performing comparisons of the \ptw spectrum above $\ptcut=7.5~\GeV$ in both the electron and muon decay channels using 8~\TeV CMS data from \cite{Khachatryan:2016nbe}. The \ptw distributions are normalised to the $\mathcal{O}(\alpha_S^2)$ inclusive result (also computed using \NNLOJET) in a similar manner to the experimental data, allowing cancellation of large systematic errors such as the experimental uncertainty in the luminosity determination. The fiducial volume is defined by the lepton cuts
\begin{align} % Condense to 2 lines?
  p_T^{e}>25~\GeV,  \quad |\eta|^{e}<2.5,  \quad\quad p_T^{\mu}>20~\GeV,  \quad |\eta|^{\mu}<2.1\mathrm{,}
\end{align}
where $e$ and $\mu$ refer to the cuts applied for electron and muon decays of the \PW respectively. The PDF set used is again the central member of the \NNPDF~ set \cite{Ball:2017nwa} with $\alpha_S(M_Z)=0.118$ for predictions at all perturbative orders. The central scale used for both $\mu_R$ and $\mu_F$ is the transverse energy of the \PW boson
\begin{equation}
  \mu_0 = E_T = \sqrt{M_{l\nu}^2+(\ptw)^2},
\end{equation}
with scale errors given by a variation of $\mu_F$ and $\mu_R$ by factors of $\{\frac{1}{2}, 2\}$ about the central value, restricting to $\frac{1}{2}\leq\mu_F/\mu_R\leq2$. In the ratios and double ratios, this restriction is generalised to an uncorrelated scale variation whilst restricting to $\frac{1}{2}\leq\mu/\mu^{\prime}\leq2$ between all pairs of scales.

Predictions for the total \PW production rates in both the muon and electron decay modes are shown in Fig. \ref{fig:CMS_8TeV_ptw}, where we observe good agreement with data across the whole of the \ptw spectrum. In particular, we observe a shape correction in the electron channel with respect to NLO in the region $25\GeV \lesssim \ptw \lesssim 100\GeV$ that describes the data particularly well. As expected there is also a considerable reduction in the scale uncertainty bands, allowing the theory prediction to become competitive with experimental precision.

\begin{figure}[!t]
  \centering
  \begin{tabular}{@{}cc@{}}
  \includegraphics[scale=0.85]{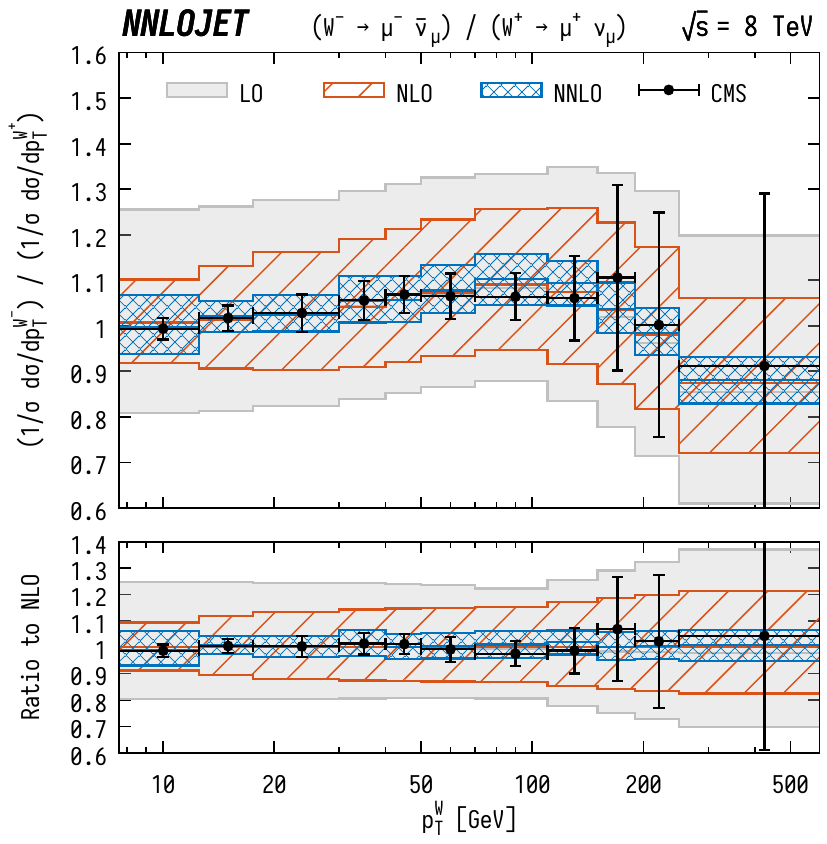} &
  \end{tabular}
  \caption{The normalised ratio of \ptw distributions $\PWm/\PWp$ in the muon decay channel. Predictions at  LO  (gray  fill),  NLO  (orange  hatched),  and  NNLO  (blue cross-hatched)  are  compared  to  CMS  data at 8~\TeV  from  \cite{Khachatryan:2016nbe}. The bands correspond to scale uncertainties estimated as described in the main text.}
  \label{fig:CMS_8TeV_ptw_ratio}
\end{figure}

Ratios of the \ptw spectra for the different \PWpm bosons also allow for further flavour specific information to be extracted from the data, of particular relevance to the $u/d$ valence quark contents in proton PDFs. EW Sudakov logarithms present at high \ptw also largely cancel in the ratios as well as many experimental systematics, which also allow for precise comparisons to data. The \ptw distribution for the ratio $\PWm/\PWp$ is shown in Fig. \ref{fig:CMS_8TeV_ptw_ratio}. Again, each of \PWm and \PWp have been normalised to their respective inclusive cross sections, such that the plot is a double ratio. One can see that there is very accurate modelling of the lineshape, which demonstrates that the PDFs capture the distribution of valence quark flavours well in this kinematic region.

%% Many calculations for inclusive cross section - DYNNLO, FEWZ, VRAP, MCFM-8.0, MATRIX, NNLOJET, but only NLO in \ptw. Can leverage \PWJ calculation for finite \ptw.
%% CUTS/SETUP
%% SCALES
%% be careful about what order alpha s and LO is
%% %\includegraphics[scale=0.8]{pdfs/W_el_ptnorm.pdf}
%% RATIO, MOTIVATION
%% EW Sudakov logarithms present at high \ptw largely cancel in ratios of distributions. Unlike at pp̄ colliders, W + and W − are produced at different rates in pp collisions due to different u/d valence quark contents in proton PDFs.
%% Z/W ratio highly relevant in the determination of W mass
%% T
%% where p T
%% W is found indirectly through p Z

%% Gives a handle on QCD modelling uncertainty between the Z and the W Important due to reconstruction of missing energy required for W but not Z Again restrict to 12 ≤ μ/μ ′ ≤ 2 between the all pairs of scales V in (1/σ dσ/dp VT )/(1/σ dσ/dp T ) → 691pt scale variation

\section{Conclusion}
We have presented NNLO QCD corrections for both the $W$ + jet process and for the transverse momentum distribution of the \PW boson, and performed a comparison these results to CMS data taken at 8~\TeV. These results were obtained using the \NNLOJET~ parton level Monte Carlo generator using the antenna subtraction method for cancellation of infra-red singularities. We see good agreement of the NNLO results with data for a range of observables in $\PWJ$ production as well as in the transverse momentum spectrum of the \PW boson.

\acknowledgments
The authors thank Xuan Chen, Juan Cruz-Martinez, James Currie, Rhorry Gauld, Marius H\"ofer, Imre Majer, Jonathan Mo, Tom Morgan, Joao Pires and James Whitehead for useful discussions and their many contributions to the \NNLOJET~code.
This research was supported in part by the UK Science and Technology Facilities Council, by the Swiss National Science Foundation (SNF) under contracts 200020-175595, 200021-172478, and CRSII2-160814, by the Research Executive Agency (REA) of the European Union through the ERC Advanced Grant MC@NNLO (340983).

\bibliographystyle{JHEP}

\bibliography{LLproceedings}

\end{document}